\documentclass[twocolumn, APS]{revtex4}

\pdfoutput=1
\usepackage{times}
\usepackage{graphicx}
\usepackage{mathtools}

\begin{document}

\title{STM Studies of Isolated Mn$_{12}$-Ph Single Molecule Magnets}

\author{K. Reaves$^{1,2,4}$}
\email{kreaves\@physics.tamu.edu}

\author{K. Kim$^4$}

\author{K. Iwaya$^4$}

\author{T. Hitosugi$^4$}

\author{H. G. Katzgraber$^{2,5}$}

\author{H. Zhao$^3$}

\author{K. R. Dunbar$^3$}

\author{W. Teizer$^{1,2,4}$}

\affiliation{$^1$Materials Science and Engineering, Texas A\&M University, College Station, TX 77843-3003, USA\\ $^2$Department of Physics, Texas A\&M University, College Station, TX 77843-4242, USA\\$^3$Department of Chemistry, Texas A\&M University, College Station, TX 77842-3012, USA\\$^4$WPI-AIMR, Tohoku University, 2-1-1, Katahira, Aoba-ku, Sendai 980-8577, Japan\\$^5$Theoretische Physik, ETH Zurich, CH-8093 Zurich, Switzerland}
 
\begin{abstract}
We study Mn$_{12}$O$_{12}$(C$_6$H$_5$COO)$_{16}$(H$_2$O)$_4$ (Mn$_{12}$-Ph) single-molecule magnets on highly ordered pyrolytic graphite (HOPG) using low temperature scanning tunneling microscopy (LT-STM) experiments. We report Mn$_{12}$-Ph in isolation, resembling single molecules with metallic core atoms and organic outer ligands. The local tunneling current observed within the molecular structure shows a strong bias voltage dependency, which is distinct from that of the HOPG surface.  Further, evidence of internal inhomogeneity in the local density of states has been observed with high spatial resolution, and this inhomogeneity appears to be due to localized metallic behavior.  These results facilitate magneto-metric studies of single molecule magnets in isolation.  As compared to bulk crystal studies, our experiments allow the specific investigation of atomic sites in the molecule.\\ 

This is a preprint of an article submitted for consideration in \href{http://www.worldscientific.com/worldscinet/spin}{Spin} \copyright$\,$2012 World Scientific Publishing Company.
\end{abstract}

\maketitle

\section{Introduction}
The Mn$_{12}$O$_{12}$(C$_6$H$_5$COO)$_{16}$(H$_2$O)$_4$ (Mn$_{12}$-Ph) single-molecule magnet consists of twelve manganese ions, four Mn$^{4+}$ ions and eight Mn$^{3+}$ ions, which are ferro/anti-ferro-magnetically coupled via oxygen bridges, forming a large net spin ($S = 10$ in the ground state).  Sixteen phenyl ligands surround the core, determining the intrinsic minimal spacing between the magnetic molecules, thus suppressing the magnetic interaction between them$\!$\cite{{Sessoli93, Saywell11}}.  This class of molecules is attractive for fundamental studies as well as for device applications because of the intriguing magnetic characteristics, the weak intermolecular magnetic interaction, and the expected electronic response in the presence of an applied magnetic field$\!$\cite{Wernsdorfer10}. The more commonly studied member of the Mn$_{12}$ family has acetate ligands$\!$\cite{Sessoli93}.  We are purposefully using phenol ligand substituted molecules to facilitate electronic contact with the highly ordered pyrolytic graphite (HOPG) substrate.\\

\begin{figure}
	\centering
	\includegraphics[width=0.5\textwidth]{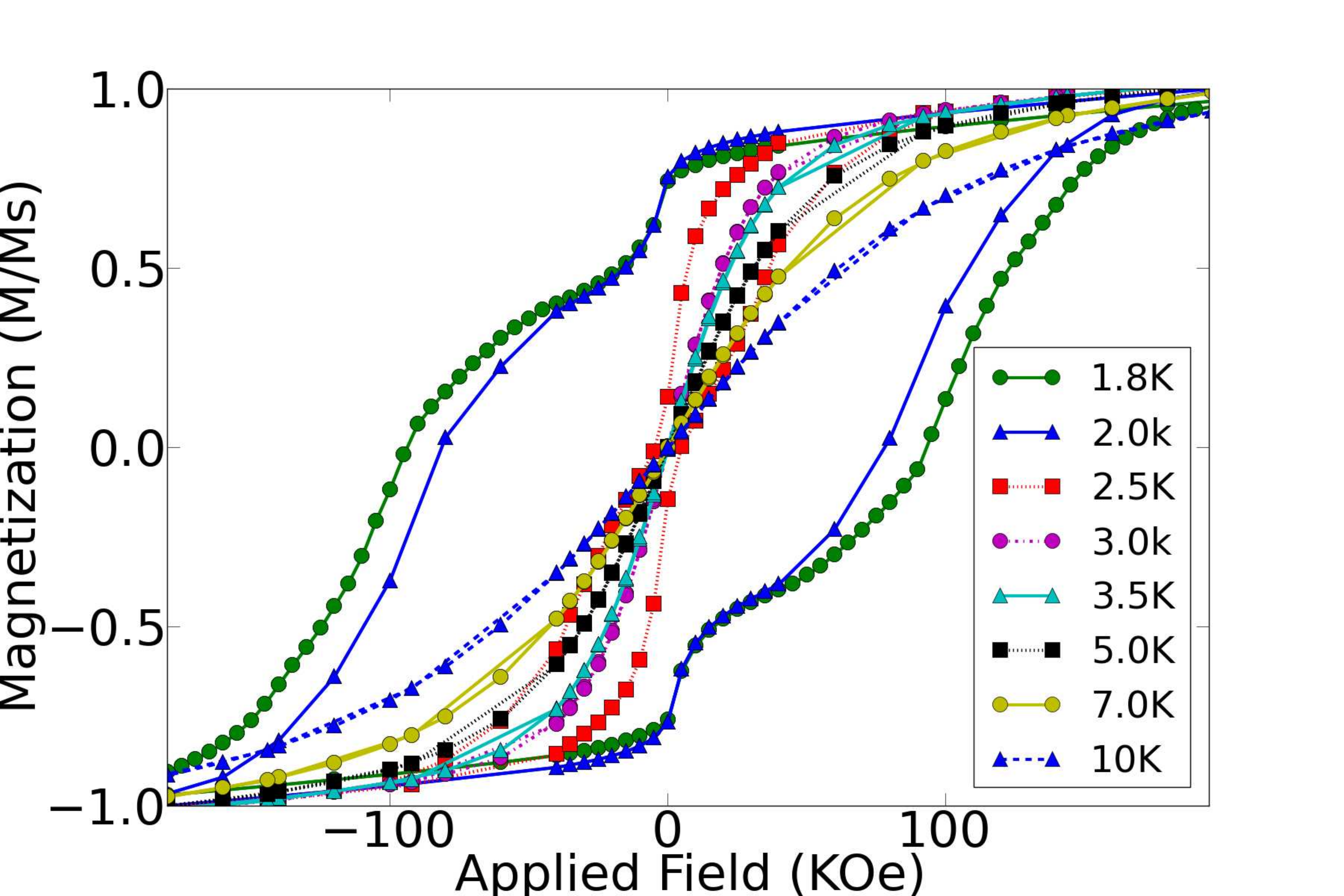}
	\caption{Magnetization of Mn$_{12}$-Ph powder sample as a function of applied magnetic field at different temperatures, $1.8\,K < T < 10\,K$.  The measured magnetization have been normalized to the saturation value ($M_s \approx 0.22\,AM^2$) to remove sample size dependence but the data has not been renormalized to remove the effect of the polymer sample container.} 
	\label{fig1}
\end{figure}

Addressing individual magnetic molecules without damaging their structure and functionality is an essential prerequisite for true molecular-level studies and device development$\!$\cite{Gatteschi09}. However, in spite of the substantial efforts to organize magnetic molecular materials and to increase the sensitivity of measurement techniques, it is not established if individual molecular units isolated on a surface are unaltered from their bulk configuration$\!$\cite{Domingo12}. Recently, scanning tunneling microscopy (STM) has been successfully utilized to identify several supramolecular metal ion assemblies deposited on HOPG surfaces at room temperature. The metal centers in the assemblies, such as Fe, Co, Cu and Mn, are recognizable as evident `bright' spots in two-dimensional scanning tunneling spectroscopic (STS) images$\!$\cite{Domingo12}. This is because of significantly enhanced tunneling currents increasing near the metal centers. The STS image-based approach provides the spatial location and arrangement of the core elements of the molecule under the STM tip which is a crucial clue to confirming the internal electronic structure of the molecule dynamically while also observing the arrangement of the local topological configuration with atomic resolution. The latter will be an important feature, particularly in any spin-polarized tunneling studies, to understanding the magnetic connectivity of the spin network within the magnetic cluster. In this paper, we present the STS image-based approach to identify Mn$_{12}$-Ph single-molecule magnets deposited on a HOPG surface, by a solution evaporation technique, and a high resolution low temperature scanning tunneling microscopy (LT-STM) based approach to study the interior structure of the Mn$_{12}$-Ph.

\section{Experimental Procedure}
In order to prepare a sample for the STM study, concentrated Mn$_{12}$-Ph solutions were created by dissolving the original Mn$_{12}$-Ph powder material in benzene.  Various concentrations were studied ($\sim\!0.02\,\,to \sim\!200\,mM$)  in an effort to have a dense enough surface to assure observation of molecules with the STM tip  after a short search but still dilute enough to obtain significant numbers of individual molecules in isolation on the surface.  Magnetization curves (DC Hysteresis) of micro-crystalline powder (non-solvated) were measured as a function of applied magnetic fields ($|H|(T)\le6\,T$ for $1.8\,K < T < 10$\,K) using a Quantum Design SQUID magnetometer and were taken across a range of temperatures (\textit{Fig.\,1}).   Samples were prepared for SQUID magnetometry by measuring small amounts of freshly produced micro-crystalline powder of Mn$_{12}$-Ph and placing them in a sample container affixed to a cold finger for insertion into the SQUID. Since the magnetization data for the graph (vertical axis) is dependent on the sample mass, dividing by the saturation magnetization measurement  yields a unit-less result which is sample size and mass independent, allowing for easier comparison to other data for different samples and similar molecules (i.e. Mn$_{12}$-Ac\cite{Sessoli93}).  There is a known aberration in the graph due to a polymer sample container, which has not been subtracted from the data presented.  In the blocked state, below 3.0K, steps in the hysteresis curves were observed, as expected$\!$\cite{{Friedman96, Friedman99}} .\\ 

\begin{figure}
	\centering
	\includegraphics[width=0.5\textwidth]{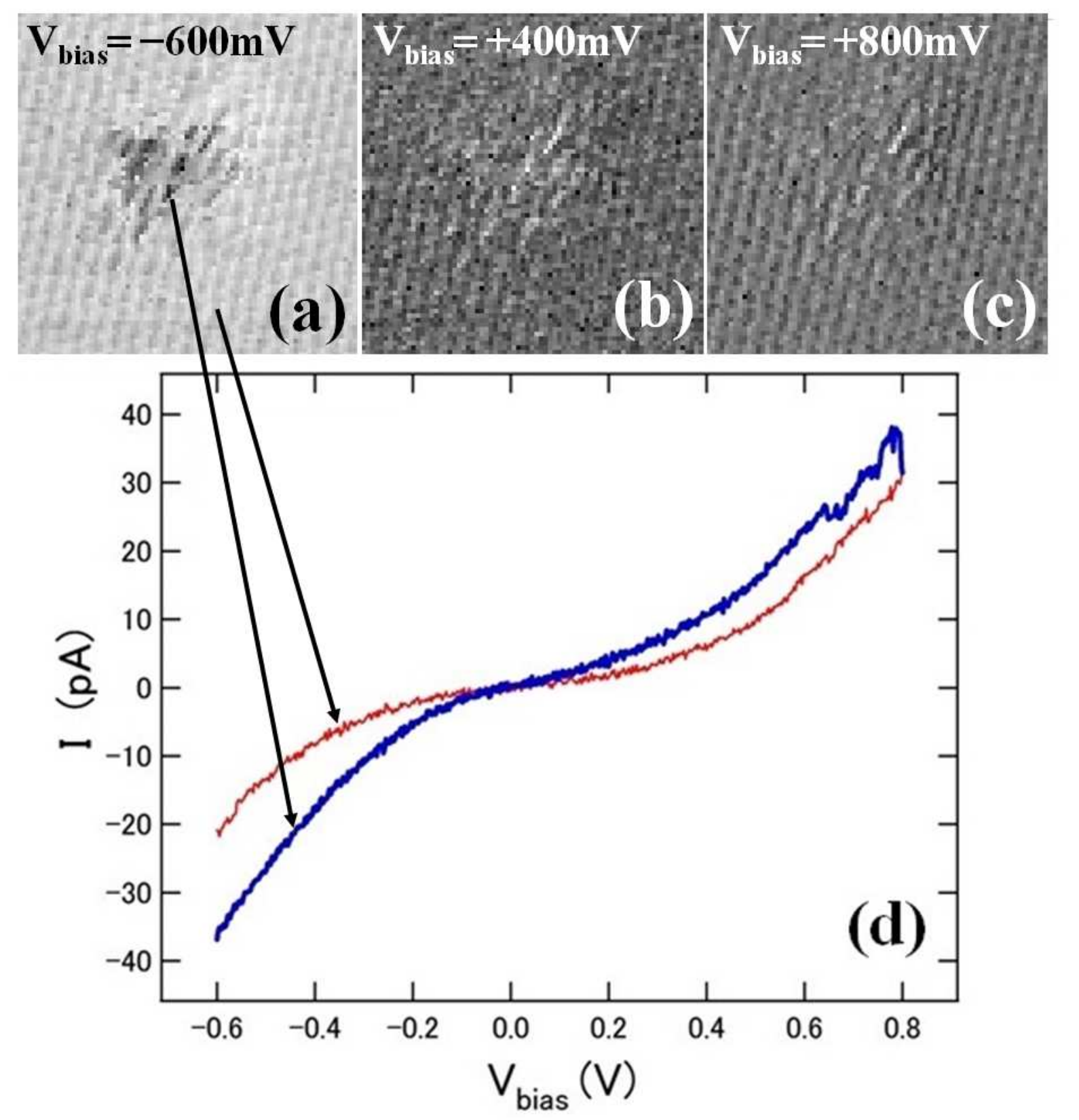}
	\caption{Two-dimensional current mapping obtained at a location including an isolated object at three different bias voltages, $-600\,mV$ (a), $+400\,mV$ (b), $+800\,mV$ (c) Tunneling current ($pA$) with respect to bias voltage ($V$) measured at  two locations marked by arrows (d).  Scan size: $4.5\times4.5\,nm^2$. STM chamber temperature: $\sim\!4.9\,K$.}
	\label{fig2}
\end{figure}

Samples were produced by dropping a small ($\sim\!1\,\mu$L) droplet of $\sim\!0.2 mM$ solution of Mn$_{12}$-Ph in Benzene via pipette onto freshly cleaved (in ambient conditions) HOPG substrates.  After drying, the samples were loaded into  the LT-STM system for observation at $\sim\!80\,K$ and $\sim\!4.9\,K$.  Once a molecule was located on the surface the tunneling conditions were altered to acquire, in the range of $\pm1\,V$, tunneling current vs. voltage (I-V) data for each point in a $128\times128$ grid in a given topographic region near several molecules at $\sim\!4.9\,K$ (\textit{Fig.\,2}).  In addition, manually altering the tunneling conditions and then taking STS data of only selected points assured that both broad trends and specific behaviors were studied. \\

Typical experiments reported here using high resolution LT-STM to resolve structures internal to the molecule perimeter used the sample preparation method described above.  \textit{Fig.\,3} shows a representative result from these studies.  This method of sample preparation, however, did not provide homogeneous films of isolated molecules in significant density.  The sample preparation method was therefore altered in an effort to produce samples with more consistent and denser coverage of isolated molecules using a common technique$\!$\cite{Moroni08}.  For this, a $\sim100\,mM $of Mn$_{12}$-Ph in benzene solution was sprayed into vacuum ($\sim$10$^{-7}$ torr) to selectively deposit the Mn$_{12}$-Ph onto HOPG which had been freshly cleaved.  The solvent which has a significantly lower molecular weight that the Mn$_{12}$-Ph molecule is preferentially pumped away during the deposition process.  High resolution STM images of  Mn$_{12}$-Ph molecules on the surface show distinct internal features which appear qualitatively similar across different locations and on different samples.  In \textit{Fig.\,4} the XY resolution is 0.195 nm per pixel and the total Z  between white and black covers a range of $1.99\,nm$. 

\begin{figure}
	\centering
	\includegraphics[width=0.5\textwidth]{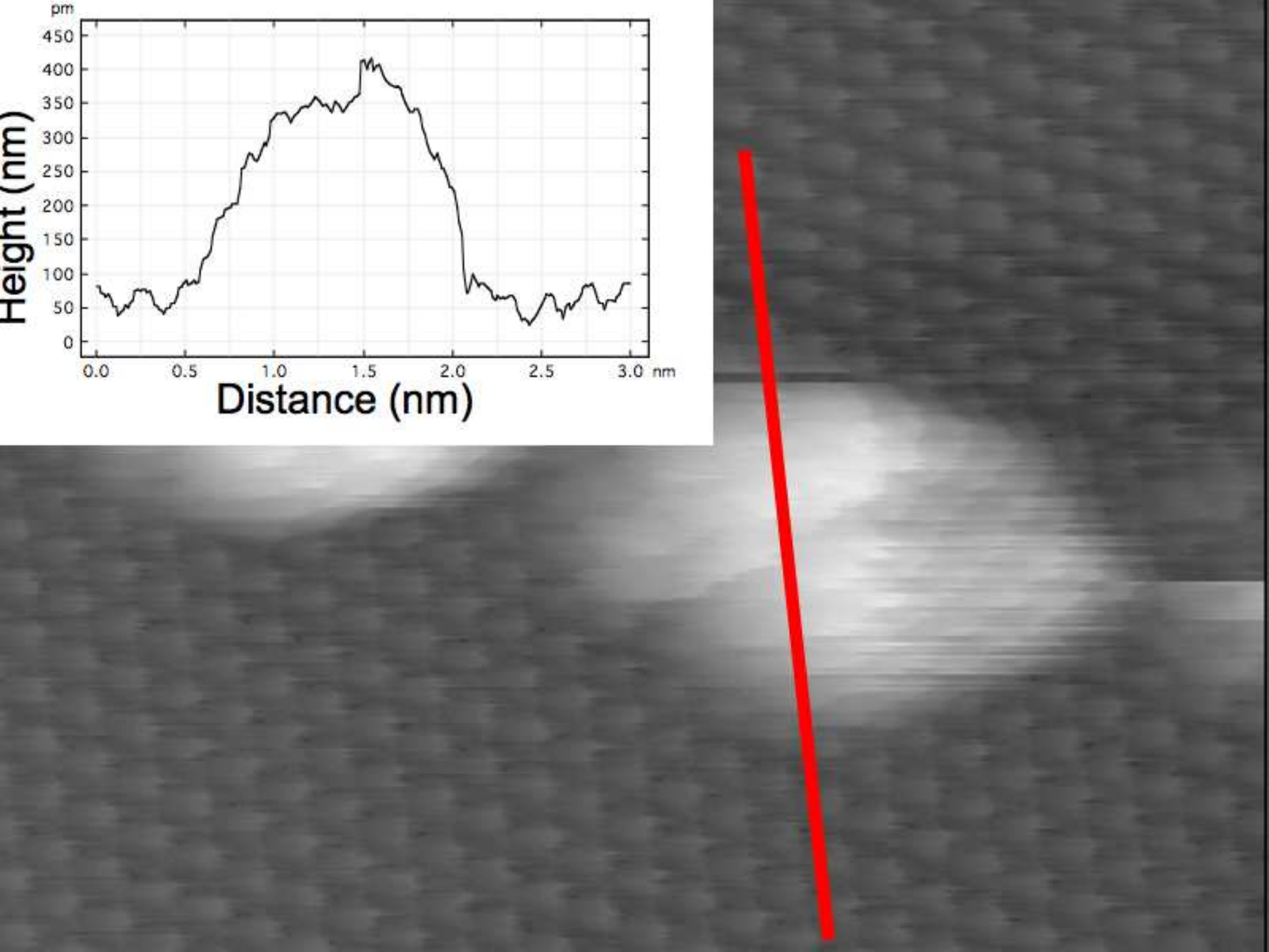}
	\caption{An STM topography image of molecular objects isolated on a HOPG surface at $\sim4.9 \,K and\,\,V_B = -500\,mV$. Inset: Height profile with respect to displacement along the vertical (red) line. The horizontal (black) line is a scanning artifact.}
	\label{fig3}
\end{figure}

\begin{figure}
	\centering
	\includegraphics[width=0.5\textwidth]{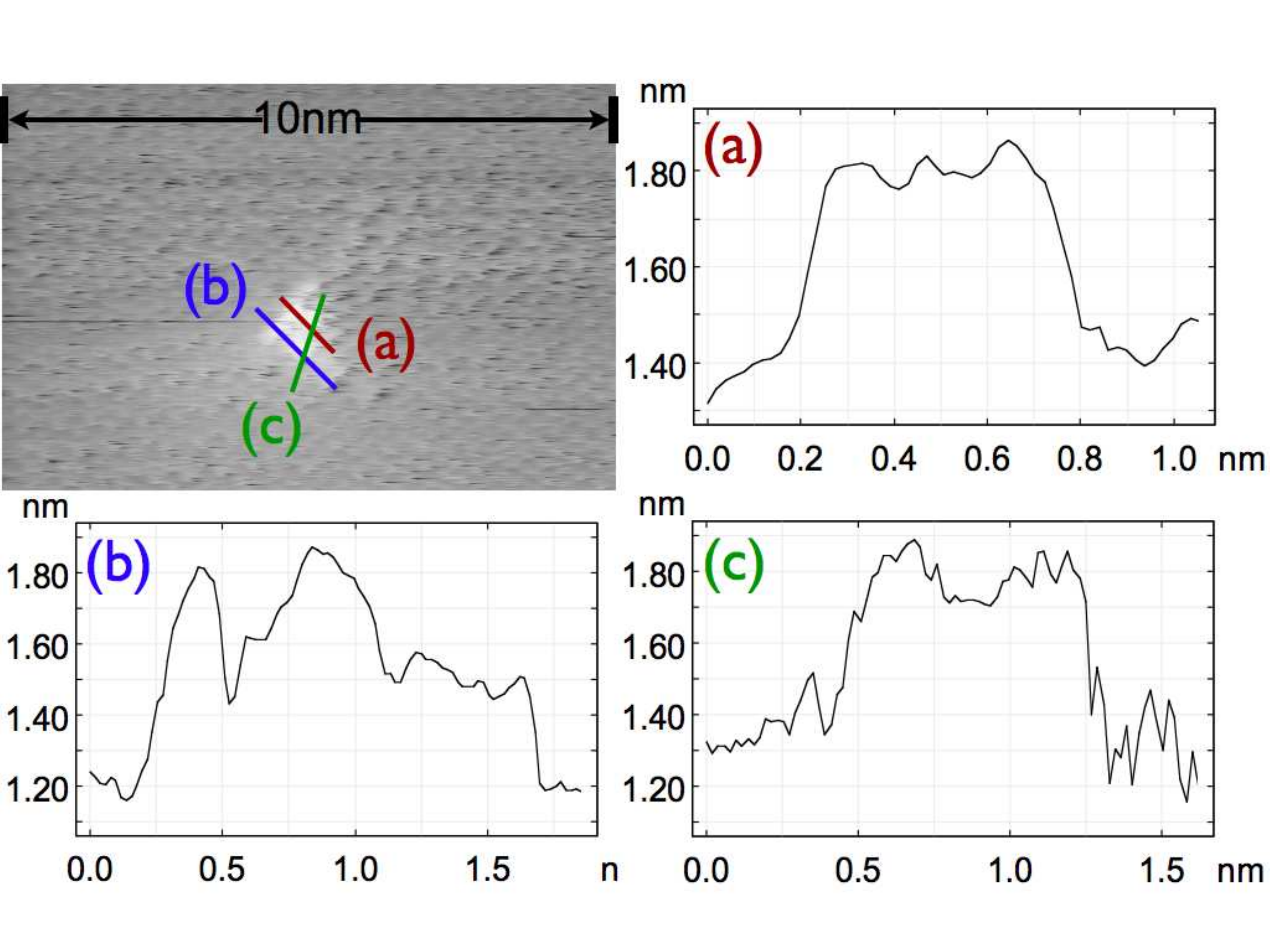}
	\caption{Topographic image is from a 10nm$\times$10nm STM image at $\sim$80\,K of a molecule like feature with panels (a), (b), and (c) overlaid over part of the topographic image.  Subfigures (a), (b), and (c) are height profile plots with respect to displacement along the indicated lines.}
	\label{fig4}
\end{figure} 

\section{Results}
We compare our STM results to those in the literature for related molecules in the Mn$_{12}$ family and have quantitatively (size, general profile, and bias dependence) similar results$\!$\cite{Saywell11} in our topographic images, although we are comparing related (not identical) molecules.  We achieve lower densities of isolated molecules on the surface than have been reported using electro-spray deposition or solution spray deposition$\!$\cite{{Kahle12, Moroni08}} .The care taken with the surface and solvated molecule solution introduced to the surface implies that the observed objects are either related to the solvent or the solvate (or some combination of both).  The size of the objects coupled with the local density of states (LDOS) behavior of the interior of the molecules$\!$\cite{Domingo12} indicates that the molecule-like objects are metal-centered large molecules.  Control experiments were performed using clean (freshly cleaved) HOPG substrates with no solvent or molecules deposited, and also HOPG substrates with only solvent (benzene) added, for each sample preparation method.  No comparably-sized molecule-like objects were observed in the control experiments.  The bias evolution dependent behavior indicated that the molecules are physisorbed to the surface, as expected.  Coupled with the control experiments this indicates that the objects are not surface defects.\\ 

We have evaluated the results of our bias evolution experiments. Characteristics of Mn$_{12}$-Ph on HOPG show distinctly different I-V behavior when compared to the graphite background.  As can be seen in \textit{Fig.\,2}, the I-V behavior for what are nominally the same XY locations are distinct at different bias voltages and diverge from the behavior of locations on the HOPG substrate.  We observe several distinct behaviors in the I-V curves from locations within the molecule's perimeter via our STS studies.  None of the internal features show a purely metallic behavior, but all of the points inside the perimeter of the molecule behaved more metallicaly, when compared to the HOPG background locations.  These findings are consistent with the behavior of similar organo-metallic cluster molecules$\!$\cite{Domingo12}.  The observed bias dependence for these molecules implies that they are physisorbed to the surface which leads to difficulties in maintaining consistent tip conditions during imaging.  This difficulty in maintaining stable tunneling conditions explains scanning artifacts like the horizontal line in \textit{Fig.\,3}.  Further, we notice that not all portions of the molecule have the same bias dependence across the range of evolution; this is consistent with the reported behavior of other organo-metallic cluster molecules$\!$\cite{Domingo12}.\\

STS data are comparatively low resolution compared to the LT-STM images.  This is due to the length of time required to perform complete bias spectra at each point, even in a restricted range ($\pm1\,V$).  Several spectra must be taken for each point to arrive at an average.  This long time period requires a correspondingly lower spacial resolution.  Our high resolution LT-STM images are 512$\times$512 pixels regardless of the scale of the image.  Similarly, our STS images are 128$\times$128 pixels while covering the same spatial area.  Some of the inhomogeneity of the interior of the molecules can be lost to this pixelation, but the technique returns complete bias spectra for each point in the image.  Conversely, high resolution STM images can be taken much more quickly (which limits how much the tunneling conditions can change in any single image frame) but only capture the behavior at a single bias voltage for every image.  This makes it impractical to simulate STS through altering the bias voltage of the tunnel junction to complete our $\pm1\,V$ bias range due to the similar time constraints as using STS with higher image resolution.  STS can be done for specific points, and STM can be taken at or near particular bias voltages to explore the behavior of the Mn$_{12}$-Ph on the surface.  The are alternative implementations of the same basic tunneling experiment.  By focusing each implementation where it is most useful we can receive complimentary results.  Within the high resolution STM topographic images, we see distinct clusters within the outer perimeter of the molecule.  These clusters appear in distinct geometries which can be compared to DFT simulations and models of landing configurations.  The qualitative similarity of the location and conformation of these clusters across different molecules on the sample surfaces indicates that we are observing the metallic atoms in the core of the Mn$_{12}$-Ph molecule.  We expect to improve resolution as we progress to lower temperatures.  At $\sim80\,K$ we think we already have sufficient XY resolution to distinguish metallic atoms (namely Mn) of Mn$_{12}$-Ph, and we are working on modeling of landing orientations and DFT simulations to compare to.

\section{Discussion}
Mn$_{12}$-Ph is not the most commonly studied from the Mn$_{12}$ family of molecules, but the relative ease of changing functional (or inert) organic ligand groups is an added incentive for studying the family of molecules for later adaptation to application.  The two experimental approaches described, while explained separately, are complimentary and serve to explore the electronic structure of isolated Mn$_{12}$-Ph on HOPG while being capable of observing the interaction of the molecule with the substrate. We have shown Mn$_{12}$-Ph in isolation and the results resemble those reported for single molecules with metallic core atoms and organic outer ligands$\!$\cite{Domingo12}. The local tunneling current observed within the molecular structure shows a strong bias voltage dependency, which is distinct from that of the HOPG surface in its inhomogeneity as well as its tendency towards more metallic behavior.  The I-V curves from different locations have been studied inside the molecule as well as outside the molecule in locations corresponding to the substrate and with different tunneling conditions (temperature, tunneling current, bias voltage).  Evidence of internal inhomogeneity in the LDOS has been observed with high resolution and is likely the result of localized metallic behavior (due to the Mn atoms).

\section{Conclusion}
We have shown that through two different sample preparation methods we can study isolated Mn$_{12}$-Ph on HOPG via LT-STM.  This work consists of two main thrusts, bias evolution STM studies and STS studies.  STS allows us to have I-V data for every point in our topographic image at the expense of limiting topographic resolution.  A systematic study of the I-V behavior of background locations compared to locations interior to the perimeter of Mn$_{12}$-Ph is ongoing.\\

Combining the two related experimental techniques (STM and STS) allows us to study localized behaviors as well as compare the overall behavior of different molecules in a broader context.  By studying both the internal inhomogeneity in the LDOS as well as the larger (physical) scale of the molecule compared to the substrate we can study the interaction with the surface as well as the internal interactions of the molecule during tunneling events.  The studies reported here are at $\sim80\,K$ and $\sim4.9\,K$, while the planned operating conditions are at lower temperatures (below the blocking temperature, $\sim3\,K$), and will explore the magnetic dependence of the LDOS.  We expect better resolution of internal features at lower temperatures, and to observe spin dependent tunneling events at the molecular scale.  With resolution of individual manganese atoms within the molecule we will for the first time be able to compare with our DFT simulations as well as the reported phenomena in larger crystals$\!$\cite{{Friedman96, Friedman99}}.

\section{Acknowledgments}
We would like to acknowledge the WPI-AIMR institute for support.  We would also like to acknowledge the National Science Foundation for an equipment grant to purchase a SQUID magnetometer (NSF-9974899) and Andrew Prosivirin for technical assistance.  We would like to thank K. Park for insightful discussions.  Helmut G. Katgraber would like to acknowledge support from the SNF (Grant No.~PP002-114713) and the NSF (Grant No.~DMR-1151387).

\end{document}